# Astro2020 Science White Paper

# The Future of Exoplanet Direct Detection

**Thematic Areas:** ✅ Planetary Systems ☐ Star and Planet Formation
☐ Formation and Evolution of Compact Objects ☐ Cosmology and Fundamental Physics
☐ Stars and Stellar Evolution ☐ Resolved Stellar Populations and their Environments
☐ Galaxy Evolution ☐ Multi-Messenger Astronomy and Astrophysics

**Principal Author:**
Name: John Monnier
Institution: University of Michigan
Email: monnier@umich.edu
Phone: 734-763-522

**Co-signers:** (names and institutions)
Gioia Rau — NASA/GSFC
Ellyn K. Baines — Naval Research Lab.
Joel Sanchez-Bermudez — Instituto de Astronomía de la UNAM
Martin Elvis — Center for Astrophysics | Harvard & Smithsonian
Sam Ragland — W.M. Keck Observatory
Rachel Akeson — Caltech/IPAC
Gerard van Belle — Lowell Observatory
Ryan Norris — Georgia State University
Kathryn Gordon — Agnes Scott College
Denis Defrère — University of Liège
Stephen Ridgway — NOAO
Jean-Baptiste Le Bouquin — University of Grenoble, University of Michigan
Narsireddy Anugu — University of Exeter
Nicholas Scott — NASA Ames
Stephen Kane — University of California, Riverside
Noel Richardson — University of Toledo
Zsolt Regaly — Konkoly Observatory, Research Center for Astronomy and Earth Sciences, Budapest, Hungary
Zhaohuan Zhu — University of Nevada, Las Vegas
Andrea Chiavassa — Université Côte d'Azur, Observatoire de la Côte d'Azur, CNRS, Lagrange, CS 34229, Nice, France
Gautam Vasisht — JPL-Caltech
Keivan G. Stassun — Vanderbilt University
Chuanfei Dong — Princeton University

Olivier Absil	University of Liege
Sylvestre Lacour	Observatoire de Paris
Gerd Weigelt	Max Planck Institute for Radio Astronomy
Douglas Gies	Georgia State University
Fred C Adams	University of Michigan
Nuria Calvet	University of Michigan
Sascha P. Quanz	ETH Zurich, Switzerland
Catherine Espaillat	Boston University
Tyler Gardner	University of Michigan
Alexandra Greenbaum	University of Michigan
Rafael Millan-Gabet	Giant Magellan Telescope Organization
Chris Packham	University of Texas at San Antonio
Mario Gai	Istituto Nazionale di Astrofisica, Osservatorio Astrofisico di Torino
Quentin Kral	Paris Observatory
Jean-Philippe Berger	Université Grenoble Alpes
Hendrik Linz	MPIA Heidelberg
Lucia Klarmann	Max-Planck-Institut für Astronomie
Jaehan Bae	Carnegie Institution of Washington
Rebeca Garcia Lopez	Dublin Institute for Advanced Studies
Gallenne Alexandre	ESO (Chile)
Fabien Baron	Georgia State University
Lee Hartmann	University of Michigan
Makoto Kishimoto	Kyoto Sangyo University
Melissa McClure	University of Amsterdam
Johan Olofsson	Universidad de Valparaíso (Chile)
Chris Haniff	University of Cambridge, UK
Michael Line	Arizona State University
Romain G. Petrov	Université de la Côte d'Azur
Michael Smith	University of Kent
Christian Hummel	European Southern Observatory (ESO)
Theo ten Brummelaar	CHARA/Georgia State University
Matthew De Furio	University of Michigan
Stephen Rinehart	NASA-GSFC
David Leisawitz	NASA-GSFC
William Danchi	NASA-GSFC
Daniel Huber	University of Hawaii
Edward Wishnow	Space Sciences Lab/UC Berkeley
Denis Mourard	Observatoire de la Côte d'Azur
Benjamin Pope	New York University
Michael Ireland	ANU
Stefan Kraus	University of Exeter
Benjamin Setterholm	U. Michigan
Russel White	Georgia State University

**Abstract** (optional):
Diffraction fundamentally limits our ability to image and characterize exoplanets. Current and planned coronagraphic searches for exoplanets are making incredible strides but are fundamentally limited by the inner working angle of a few $\lambda/D$. Some crucial topics, such as demographics of exoplanets within the first 50 Myr and the infrared characterization of terrestrial planets, are beyond the reach of the single aperture angular resolution for the foreseeable future. Interferometry offers some advantages in exoplanet detection and characterization and we explore in this white paper some of the potential scientific breakthroughs possible. We demonstrate here that investments in "exoplanet interferometry" could open up new possibilities for speckle suppression through spatial coherence, a giant boost in astrometric precision for determining exoplanet orbits, ability to take a census of young giant exoplanets (clusters <50 Myr age), and an unrivaled potential for infrared nulling from space to detect terrestrial planets and search for atmospheric biomarkers. All signs point to an exciting future for exoplanets and interferometers, albeit a promise that will take decades to fulfill.

INTRODUCTION
The most exciting long-term goal of exoplanet studies is to detect and characterize true Earth analogues around nearby stars, with the hope of detecting life through atmospheric biomarkers. We are still far from achieving this goal but a roadmap exists to reach this epic achievement. Currently, transit studies have made the most progress, opening up atmospheric studies through primary transit spectroscopy and secondary eclipse measurements . Space and ground-based efforts have succeeded in characterizing atmospheres of giant exoplanets and even Neptune-like systems when very close-in to their central stars.  There may even be a chance to probe exoplanets in the Habitable Zone using JWST for candidates around M-stars (Kreidberg et al. 2018).

In order to push toward true Earth analogues around solar type stars, we must find a way to isolate the exoplanet from the host star and to suppress photon noise arising from the stellar light.  This implies the development of two technologies -- 1) we must have enough angular resolution to separate the star from planet, and 2) we must suppress the starlight to maintain the sensitivity to detect the planet.  To achieve 1) we must use the world's largest telescopes with extreme adaptive optics (AO), and 2) we must develop a highly optimized coronagraph.

The Gemini Planet Imager (GPI; Macintosh et al. 2014) and VLT/SPHERE (Beuzit et al. 2019) projects were built to apply extreme AO on the current 8m class telescopes and have surveyed hundred of stars in order to directly direct their exoplanets and to measure their spectra.  These instruments greatly improved our ability to detect exoplanets and, while final results are not yet available, it is clear that only a few exoplanets were detected (i.e., Macintosh et al. 2019). This result is telling us important information on the demographics of exoplanets beyond 20 au and their infrared cooling curves as a function of age.

**Fundamentally, we currently lack the angular resolution ("inner working angle") necessary to isolate and suppress starlight to radically increase the number of imaged exoplanets.**

Other white papers will discuss the immense potential for new single-aperture space telescopes and next-generation AO on 30m-class telescopes.   Here we wish to explore a more fundamental (and technologically daunting) breakthrough in capabilities if we could achieve an order-of-magnitude improvement in inner working angle, from both ground and space.

CURRENT STATUS and ISSUES
In this short white paper, we will briefly address four breakthroughs that could be made possible by "exoplanet interferometry:"  A) speckle suppression through spatial coherence, B) micro-arcsecond-level astrometry for detected exoplanets to determine orbits without waiting for full period, C) the ability to census young self-luminous exoplanets in star-forming regions even from the ground, and D) the ability to characterize rocky planets in the mid-infrared.

**Breakthrough A: Speckle suppression through spatial coherence**
Using GPI and SPHERE, we can now search for signs of molecular absorption (or emission) in the near infrared for detected exoplanets.  In order to extract the spectrum of an exoplanet, one must correct for contamination of stellar speckles due to residual wavefront errors. An innovative way to remove speckles from the star is through **spatial coherence.** With foreknowledge of the location of HR 8799e, the VLT interferometer used the GRAVITY instrument to inject light from the star and planet into two separate fibers of the combiner. The light from the planet had a long-baseline interference fringe at a different delay line position than the star; thus, any residual starlight that entered the 'exoplanet fiber' was not coherent at the same time as the planet's light. Figure 1 compares the GPI/SPHERE spectrum of HR8799e with a new result of using all four 8m UT telescopes with GRAVITY —the improvement is incredible (Lacour et al. 2019).

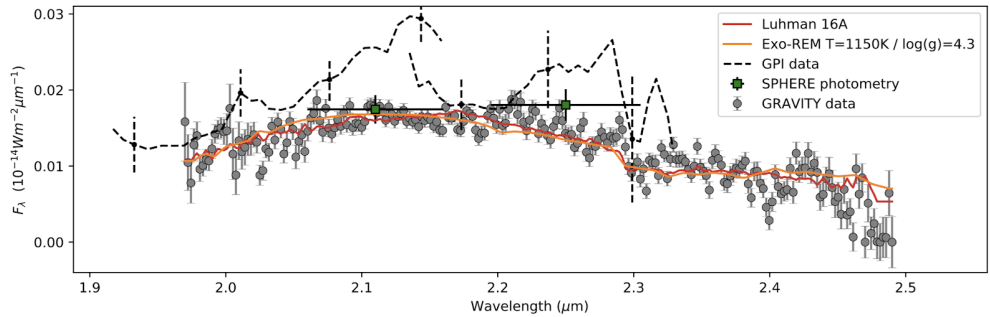

Figure 1. First demonstration of exoplanet characterization using long-baseline interferometry. Lacour et al (2019) used the VLTI-GRAVITY combiner (4x8m telescopes with baseline maximum 120m) to isolate the planet light from the contaminating stellar light via spatial coherence.

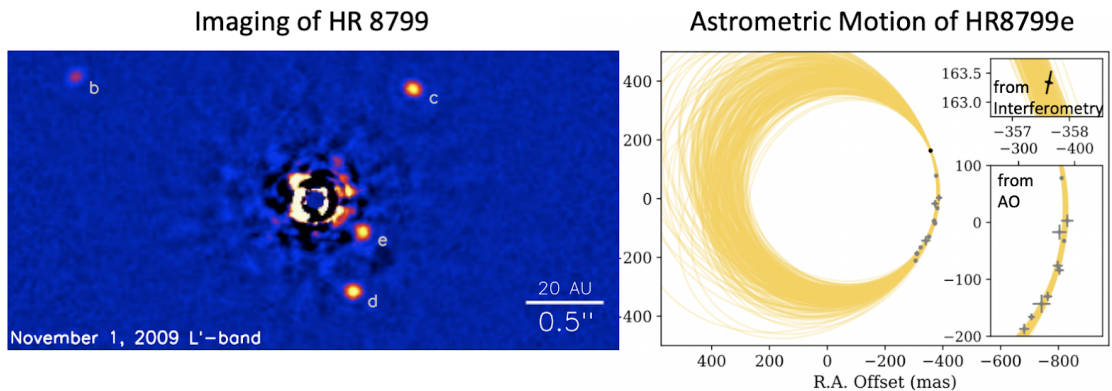

Figure 2. (left) The power of direct imaging is to detect many exoplanets at once and track their orbits (from Marois et al. 2010). (right) Here we see a summary of astrometric results for HR8799e, including the first results from VLTI-GRAVITY interferometer for HR8799e (Lacour et al. 2019). Bigger telescopes and longer baselines lead to better astrometry.

**Breakthrough B: Microarcsecond-level astrometry**
Long-baseline interferometry opens up new parameter space in precision astrometry, although historically it has been difficult to fully calibrate. The VLTI-GRAVITY combiner mentioned above was designed for astrometry of stars around the Sgr A* in the Galactic Center and has recently published impressive results (GRAVITY Consortium 2018). Such astrometry can also be used for exoplanets. Figure 2 shows an single image of the HR 8799 system and the derived motion of

HR8799e, including a new point from VLTI-GRAVITY with ≲100 microarcsecond precision. Additional data in the new few years should yield tight constraints on the many-decade orbit of this exoplanet. It is expected that additional data for all the known exoplanets can yield relative masses too (Wang et al. 2018) based on their combined gravitational effects.

FUTURE ADVANCES
While Breakthroughs A and B are partially achievable now from the ground, we will need significant new technical developments to achieve the next two Breakthroughs.

**Breakthrough C: Census of young self-luminous exoplanets in star-forming regions**
Young (<50 Myr) exoplanets would seem easy to detect; they are still warm from their formation making their infrared contrast very favorable. However, most star forming regions are >100 pc away, meaning that for a typical 0.2" inner working angle with today's telescopes, we can only detect planets beyond 20 au—a region not expected, or found to be, filled with giant exoplanets. Here, the need for higher angular resolution—not contrast—is paramount.

Figure 3 shows a simulated observation using a potential next-generation near-infrared interferometer (PFI, Planet Formation Imager), demonstrating the exciting potential to collect imaging snapshots of young giant planets with current technology from the ground. We would be able to definitively answer some of the most important questions in exoplanets and planet formation theory if we could make these kind family portraits around a few hundred young disks in Taurus, Ophiuchus, Orion, and young clusters of different ages. For instance, Figure 4 shows how the locations of giant planets are expected to evolve with time, depending on the level of migration and/or dynamical instabilities. Direct detection imaging of young exoplanets with Extremely Large Telescopes (ELTS; inner working angle ~0.07") and with interferometers (inner working angles <0.01") will explore the full region of interest from 0.1 au to 10 au.

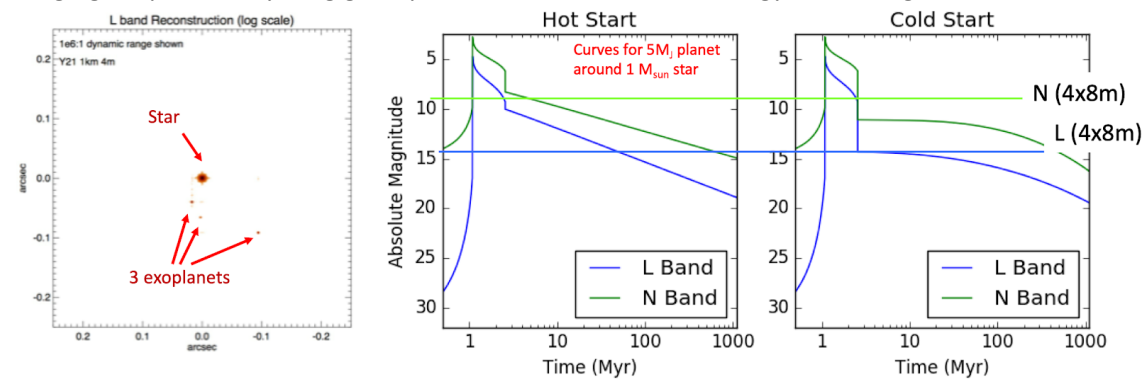

Figure 3. (left panel) Simulated observations of a young planet-forming disk (based on hydrodynamic simulation of Dong et al. 2015). Three of the four giant exoplanets were detected at L band (3.8 micron) using a ground-based 21x 4m telescope interferometer. (middle and right panels) Exoplanet detection limits for an optimized ground-based 4x 8m telescope interferometer (similar to VLTI) are shown by the colored horizontal lines (e.g., Wallace & Ireland 2019), critically depending on how the planets cool (Spiegel & Burrows 2012) and the mode of accretion (Zhu 2015). Further details can be found in Monnier et al. (2016, 2018).

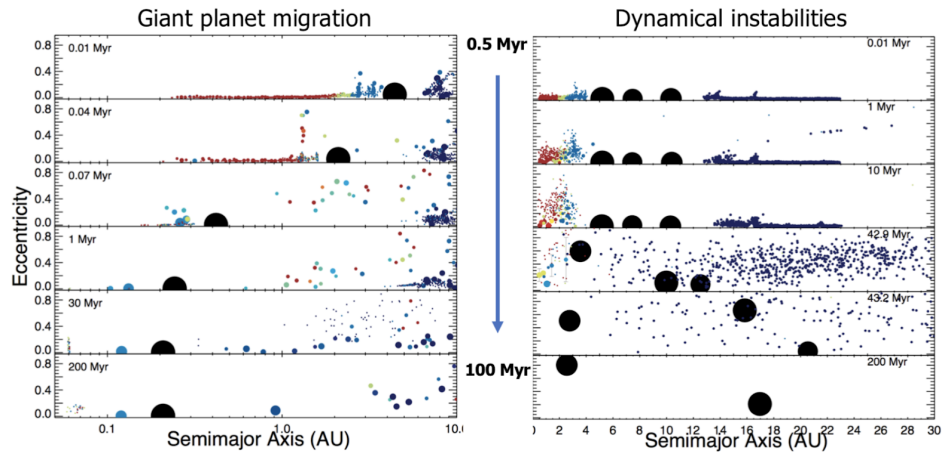

**Figure 4. The distribution of giant planets are expected to change over the first few hundred Myr. Direct detection of giant exoplanets still warm from formation could unambiguously tell us the relative importance of migration vs dynamical instabilities (from Raymond et al. 2006, 2011)**

**Breakthrough D: Ability to characterize rocky planets**
We recognize the difficulty in directly detecting thermal emission from rocky planets around even the nearest stars. There is some hope that JWST and the suite of instruments on the 30m class ELTs will be successful if the right planets exist around our nearest neighbors (e.g., ELT/METIS -- Quanz et al. 2015; TMT/MICHI -- Packham et al. 2018). There will be other white papers discussing these exciting possibilities. Here, we push for even more powerful facility architectures that could open up **hundreds more** candidates for study—one such facility is a space-based mid-infrared nulling interferometer.

By sitting on a destructive interference point, nulling interferometry allows host star light to be cancelled out while permitting planet light to interfere, thus low-mass planets can be detected without the photon noise from the host star (Bracewell 1978). NASA and ESA studied the Terrestrial Planet Finder Interferometer (TPFI, Lawson et al. 2007) and DARWIN (Cockelle et al. 2009) mission concepts about 10 years ago in a time before Kepler. Now, armed with a better knowledge of exoplanet demographics, we can make accurate mission studies of the exoplanet yield with various technologies. Figure 5 shows a recent calculation for rocky planet detections based on a 4x 2.8m telescope nulling interferometer with maximum 500 meter baselines (note this facility would have the same collecting area as JWST). The detection statistics are very favorable and would highly complement the visible light exoplanet detections from a future HabEx or LUVOIR coronagraphy mission.

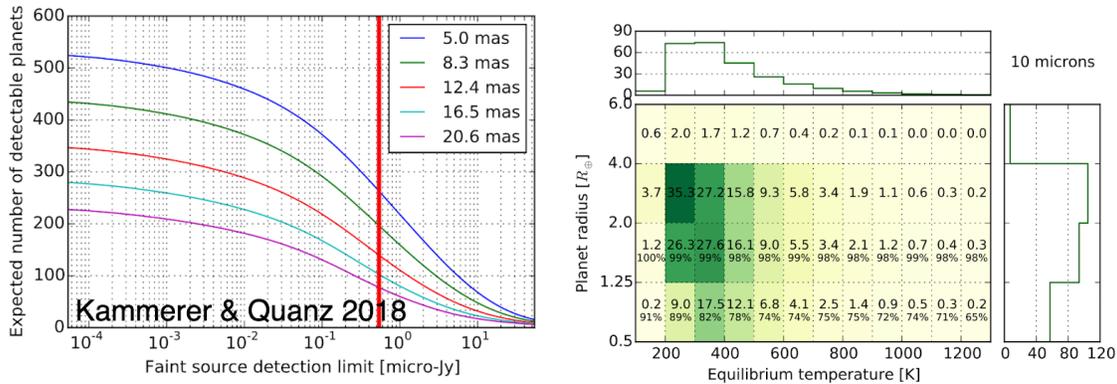

**Figure 5.** Using realistic exoplanet demographics based on the Kepler Mission, Kammerer & Quanz (2018) updated the yield for a space-based mid-IR interferometer (e.g., LIFE Mission Concept, Large Interferometer for Exoplanets). (left panel) Number of exoplanets detectable depending on the inner working angle and sensitivity. (right panel) A mid-IR interferometer could detect large numbers of rocky planets, complementing visible-light direct detection schemes being imagined for next generation space telescope missions, such as HabEx/LUVOIR.

SUMMARY
Directly imaging exoplanets is an efficient way to detect and characterize large numbers of exoplanets—once we can technically achieve a small enough working angle and the required sensitivity. The initial yield of exoplanets from GPI, SPHERE, and others is the tip of the iceberg and the ELTs will deliver a huge increase of detections, given our current understanding of exoplanet demographics.

In this white paper, we outlined some of the breakthroughs in store by combining the high angular resolution of infrared interferometry with exoplanet studies. Ground-based interferometers can currently support and enhance the characterization of known exoplanets (both in measuring spectra and astrometry) while a future NIR/MIR ground-based facility could image a large fraction of giant exoplanets in nearby star forming regions.

Furthermore, space interferometry should be re-invigorated in order to pave the way for a future mid-infrared interferometer that can measure spectra of hundreds of rocky exoplanets in a highly complementary way to the visible-light coronagraphic approaches. Space technology is quite different now compared to the time of TPFI/DARWIN; commercial development of formation flying technologies, inexpensive launch vehicles, and commodity space telescopes make space interferometry more attractive to pursue now, and the new exoplanet demographic knowledge derived from Kepler makes the scientific payoff a near certainty.

**References**

Beuzit, J.-L., Vigan, A., Mouillet, D., et al. 2019, arXiv:1902.04080

Bracewell, R.N. 1978, Nature, 274, 780

Cockell, C.S., Leger, A., Fridlund, M., et al. 2009, Astrobiology, 9, 1

Dong, R., Zhu, Z., & Whitney, B. 2015, Astrophysical Journal, 809, 93

Gravity Collaboration, Abuter, R., Amorim, A., et al. 2018, Astronomy & Astrophysics, 615, L15

Kammerer, J., & Quanz, S.P. 2018, Astronomy and Astrophysics, 609, A4

Kreidberg, L. 2018, Handbook of Exoplanets, 100

Lacour et al. 2019, Astronomy & Astrophysics, in press

Lawson, P.R., Lay, O.P., Johnston, K.J., & Beichman, C.A. 2007, NASA Technical Report N, 8

Macintosh, B., Graham, J.R., Ingraham, P., et al. 2014, Proceedings of the National Academy of Science, 111, 12661

Macintosh, B., Nielsen, E., & De Rosa, R. 2019, American Astronomical Society Meeting Abstracts #233, 233, #104.01

Marois, C., Zuckerman, B., Konopacky, Q.M., Macintosh, B., & Barman, T. 2010, Nature, 468, 1080

Monnier, J.D., Ireland, M.J., Kraus, S., et al. 2016, Proc. SPIE, 9907, 99071O

Monnier, J.D., Ireland, M., Kraus, S., et al. 2018, Proc. SPIE, 10701, 1070118

Packham, C., Honda, M., Chun, M., et al. 2018, Proc. SPIE, 10702A0.

Quanz, S.P., Crossfield, I., Meyer, M.R., Schmalzl, E., & Held, J. 2015, International Journal of Astrobiology, 14, 279

Raymond, S.N., Mandell, A.M., & Sigurdsson, S. 2006, Science, 313, 1413

Raymond, S.N., Armitage, P.J., Moro-Martin, A., et al. 2011, Astronomy & Astrophysics, 530, A62

Spiegel, D.S., & Burrows, A. 2012, Astrophysical Journal, 745, 174

Wallace, A. & Ireland, M.J, 2019, submitted

Wang, J.J., Graham, J.R., Dawson, R., et al. 2018, Astronomical Journal, 156, 192

Zhu, Z. 2015, Astrophysical Journal, 799, 16